\documentclass[12pt]{article}

\usepackage{amsmath}
\usepackage{amsfonts,amssymb}
\usepackage{latexsym}
\usepackage{graphics}

\def\d{\mbox{\rm d}}

\def\half{\mbox{$\frac{1}{2}$}}
\def\cosec{\mbox{\rm cosec}}
\title{Lie Groups and Quantum Mechanics}

\author{P.G.L. Leach\footnote{permanent address: School of Mathematical Sciences, Westville Campus,
University of KwaZulu-Natal, Durban 4000, Republic of South
Africa, email: leachp@ukzn.ac.za, leachp@math.aegean.gr} $\;\;$
and$\;\;$ M.C. Nucci\footnote{email: nucci@unipg.it}}
\date{Dipartimento di Matematica
 e Informatica, \\Universit\`a di Perugia, 06123 Perugia, Italy}

\begin{document}
\maketitle

\begin{abstract}
Mathematical modeling should present a consistent description of
physical phenomena. We illustrate an inconsistency with two
Hamiltonians -- the standard Hamiltonian and an example found in
Goldstein --   for the simple harmonic oscillator and its
quantisation. Both descriptions are rich in Lie point symmetries
and so one can calculate many Jacobi Last Multipliers and
therefore Lagrangians. The Last Multiplier provides the route to
the resolution of this problem and indicates that the great debate
about the quantisation of dissipative systems should never have
occurred.
\end{abstract}

\begin{center}
{\bf {\em Dedicated to the memory of Lev Berkovich}}
\end{center}
\section{Introduction}
The mathematical description of Quantum Mechanics is largely due
to the pioneering work of PAM Dirac who recognized the connection
between the Hamiltonian description of Classical Mechanics and the
operators he needed to describe the evolution of a quantal
system\footnote {Dirac commenced his career as an electrical
engineer and it is perhaps not surprising that his recollection of
the theory of Classical Mechanics was not quite perfect
\cite{Kragh 90}.  The story  has it that he realised there was
some connection on a Sunday afternoon, but did not have a useful
text, such as Whittaker \cite{Whittaker 44 a}, at home to validate
his memory and that it was necessary to wait until the following
morning to access the University's library.}
% Some decades later it was reported that the library was then open for 168 hours a week.}.
The essential idea was that one took the Hamiltonian and wrote it
as an operator. Unfortunately the essence of the idea contained
within is the seeds for confusion.  In the case that one had an
Hamiltonian of the form $H = \half p ^ 2+ V (q) $ in the usual
notation\footnote {One should note that Dirac referred to the
energy which happened to be described by such a classical
Hamiltonian.  It is not evident if ever he contemplated
quantisation using an Hamiltonian which did not represent the
energy.} there appeared to be no questions about the correctness
of the transition from the classical description to the quantal
description.  When the momenta and coordinates were not so
conveniently separate as in the Hamiltonian above, it was
necessary to devise some rules, such as normal ordering and the
Weyl quantisation scheme, to deal with the essential
noncommutativity of the operators.  One of the beauties of
Hamilton's description of mechanics is the invariance of his
equations of motion under canonical transformation.  In an
important paper Leon van Hove \cite {van Hove 51 a} demonstrated
that quantisation -- by whatever rule one wanted to use -- and
canonical transformation did not necessarily commute.  This poses
a serious problem.  If the description of a quantal problem is
going to depend upon the choice of coordinate system, one is at
least going to have to ensure that the coordinate system being
used is physically correct.

\strut\hfill

In this paper we propose a procedure which obviates the constraint
imposed by the conflict between consistent quantisation and the
invariance of the Hamiltonian description under canonical
transformation.  We are motivated by a desire to maintain
mathematical flexibility while at the same time being physically
correct.  It appears to us that the critical problem lies with the
various quantisation schemes.  Our proposal is to require that any
quantisation scheme preserve the Noether point symmetries of the
underlying Lagrangian.  Indeed we go somewhat further.  It is well
known that there exist the potential for an unlimited number of
Lagrangians for a given dynamical system.  These Lagrangians can
be constructed through the use of the Jacobi Last Multiplier and a
knowledge of the Lie symmetries of the underlying Newtonian
equation of motion.  To each Lagrangian there corresponds an
Hamiltonian so that in the very act of constructing the basis for
the quantal problem one is already placed on the multiple horns of
a dilemma.  We propose that the Lagrangian of choice be that which
possesses the maximal number of symmetries.

\strut\hfill

We illustrate our proposal with two representations of the
Hamiltonian of the simple harmonic oscillator.  In Section 2 we
treat the simple harmonic oscillator in its standard
representation.  In Section 3 we use an alternative Hamiltonian to
demonstrate what nonsense the usual quantisation schemes produce.
In Section 4 and subsequently we show how we obtain a consistent
description using the concepts mentioned above.

\section {The Simple Harmonic Oscillator: Part I}
The standard Hamiltonian for the simple harmonic oscillator is
\begin {equation}
H = \half\left (p ^ 2 + q ^ 2\right) \label {1.1}
\end {equation}
and the corresponding Schr\"odinger equation is
\begin {equation}
i\frac {\partial u} {\partial t} = -\half\frac {\partial ^ 2u} {\partial x ^ 2} + \half x ^ 2u. \label {1.2}
\end {equation}

Equation (\ref {1.2}) possesses the Lie point symmetries
\begin {eqnarray}
& &\Gamma_{1\pm} = \exp\left [\pm 2 it\right]\left\{\pm i\partial_t - x\partial_x \pm \left [x ^ 2+\half\right]u\partial_u\right\} \label {1.5} \\
& &\Gamma_{3} = i\partial_t \label {1.6} \\
& &\Gamma_{4\pm} = \exp\left [\pm it\right]\left\{\pm \partial_x  - xu\partial_u\right\} \label {1.3} \\
& &\Gamma_{6} = u\partial_u \label {1.4} \\
& &\Gamma_{7} = s (t,x)\partial_u, \label {1.7}
\end {eqnarray}
where $s (t,x) $ is any solution of (\ref {1.2}).

The algebra of the Lie point symmetries is $\{sl (2,R)\oplus_s W\}\oplus_s\infty A_1 $, where $W $ is the three-element Heisenberg-Weyl algebra.

{We use $\Gamma_{4\pm} $ to construct a similarity solution of (\ref {1.2}).  The associated Lagrange's system is
\begin {equation}
\frac {\d t} {0} = \frac {\d x} {\pm 1} = \frac {\d u} {- xu} \label {1.8}
\end {equation}
from which it is evident that the characteristics are $t $ and $u\exp\left [\pm \half x ^ 2\right] $.  To obtain a solution which has the correct behaviour at $\pm\infty $ we choose the characteristic with the positive sign in the exponential and set
\begin {equation}
u (t,x) = g (t)\exp\left [-\half x ^ 2\right], \label {1.9}
\end {equation}
where $g (t) $ is determined by the substitution of (\ref {1.9}) into (\ref {1.2}).  It is a simple calculation to show that
\begin {equation}
g (t) = \exp\left [-\half it\right] \label {1.10}
\end {equation}
up to a multiplicative constant which we ignore.  Consequently we
have the ground-state solution
\begin {equation}
u_0 (t,x) = \exp\left [-\half it -\half x ^ 2\right] \label {1.11}
\end {equation}
corresponding to the symmetry $\Gamma_{4+} $.

Since (\ref {1.11}) is a solution of (\ref {1.2}), we may use it in $\Gamma_7 $.  Then
\begin {eqnarray}
\left [\Gamma_{4-},\,\Gamma_{7}\right]_{LB} & = & \left [\exp\left [-it\right]\left\{-\partial_x  - xu\partial_u\right\} ,\,\exp\left [-\half it -\half x ^ 2\right] \partial_u\right]_{LB} \nonumber \\
&=&   2x\exp\left [-\mbox {$\frac {3} {2} $} it -\half x ^ 2\right] \partial_u \nonumber
\end {eqnarray}
and so we have obtained another solution, namely
\begin {equation}
u_1 (t,x) = 2x\exp\left [-\mbox {$\frac {3} {2} $} it -\half x ^ 2\right]. \label {1.12}
\end {equation}
Further solutions are constructed in a similar fashion.

The symmetry, $\Gamma_3 $, acts as an eigenvalue operator since
\begin {eqnarray}
\Gamma_3 u_0 & = & \half u_0 \nonumber \\
\Gamma_3 u_1 & = & \mbox {$\frac {3} {2} $}u_1 \nonumber
\end {eqnarray}
{\it etc}.

\section {The Simple Harmonic Oscillator: Part II}

One of the attractive features of Hamiltonian Mechanics is the
preservation of the structure of Hamilton's equations of motion
under canonical transformation.  In his well-known text the
unfortunately late Herbert Goldstein presents an alternative
Hamiltonian for the simple harmonic oscillator as \cite {Goldstein
80 a} [ex 18, p 433]
\begin {equation}
H = \half\left (\frac {1} {q ^ 2} + p ^ 2q ^ 4\right). \label {2.1}
\end {equation}
The canonical transformation between (\ref{1.1}) and (\ref{2.1}) is
\begin{equation}
\tilde{q} = -\frac{1}{q} \qquad \tilde{p} = pq^2. \label{2.1a}
\end{equation}
We have a choice of methods to obtain the Schr\"odinger equation
corresponding to the Hamiltonian (\ref{2.1}).

If we use the normal-ordering method \cite{BjorD}, \cite
{Louisell} for the product involving the two canonical variables,
 the Schr\"odinger equation is
\begin {equation}
2i\frac {\partial u} {\partial t} = -x ^ 4\frac {\partial ^ 2u} {\partial x ^ 2} - 4x^3\frac {\partial u} {\partial x} + \left (\frac{1}{x ^ 2} - 6x^2\right)u. \label {2.2}
\end {equation}
Equation (\ref {2.2}) possesses the Lie point symmetries
\begin {eqnarray}
& &\Phi_{1\pm} = \exp\left [\pm 2 it\right]\left\{\pm i\partial_t + x\partial_x +\left [-\half \pm \frac {1} {x ^ 2}\right]u\partial_u\right\} \nonumber \\
& &\Phi_{3} = \partial_t \nonumber \\
& &\Phi_{4} = u\partial_u \label {2.3} \\
& &\Phi_{5} = s (t,x)\partial_u, \nonumber
\end {eqnarray}
where $s (t,x) $ is a solution of (\ref {2.2}).

If we use the Weyl quantisation scheme \cite{Weyl}, we obtain
\begin {equation}
2i\frac {\partial u} {\partial t} = -x ^ 4\frac {\partial ^ 2u} {\partial x ^ 2} - 4x^3\frac {\partial u} {\partial x} + \left (\frac{1}{x ^ 2} - 3x^2\right)u \label {2.4}
\end {equation}
and the Lie point symmetries are
\begin {eqnarray}
& &\Sigma_{1\pm} =  \exp\left [\pm 2 it\right]\left\{\pm i\partial_t + x\partial_x +\left [-\half \pm \frac {1} {x ^ 2}\right]u\partial_u\right\} \nonumber \\
& &\Sigma_{3} = \partial_t \nonumber \\
& &\Sigma_{4} = u\partial_u \label {2.5} \\
& &\Sigma_{5} = s (t,x)\partial_u, \nonumber
\end {eqnarray}
where now $s (t,x) $ is a solution of (\ref {2.4}).

Thirdly, if we use the method proposed in \cite{Leach 06 a}, the Schr\"odinger equation is
\begin {equation}
2i\frac {\partial u} {\partial t} = -x ^ 4\frac {\partial ^ 2u} {\partial x ^ 2} - 4x^3\frac {\partial u} {\partial x} + \left (\frac{1}{x ^ 2} - 2x^2\right)u \label {2.6}
\end {equation}
which has the Lie point symmetries
\begin {eqnarray}
& &\Delta_{1\pm} = \exp [\pm 2 it]\left\{\pm i\partial_t+ x\partial_x+\left [-\half \pm \displaystyle {\frac {1} {x ^ 2}}\right]u\partial_u\right\} \nonumber \\
& &\Delta_{3} = i\partial_t \nonumber \\
& &\Delta_{4\pm} = \exp [\pm it]\left\{x^2\partial_x  + \left [-x \pm\frac {1} {x}\right]u\partial_u\right\} \label {2.7} \\
& &\Delta_{6} = u\partial_u \nonumber \\
& &\Delta_{7} = s (t,x)\partial_u \nonumber
\end {eqnarray}
and now $s (t,x) $ is a solution of (\ref {2.6}).

There does seem to be something of a divergence!

In principle we can use the symmetries in (\ref{2.3}), (\ref{2.5})
and (\ref{2.7}) to construct solutions of the respective
Schr\"odinger equations and obtain the eigenvalues just as we did
for the Schr\"odinger equation (\ref{1.2}).  In the case of
(\ref{2.6}) we obtain
\begin{eqnarray}
u_0(t,x) =  x^{-1}\exp\left[-\half it - \displaystyle{\frac{1}{2x^2}}\right]
    &\quad& E_0 =  \half.      \label{2.8}
\end{eqnarray}
The results for (\ref{2.2}) and (\ref{2.4}) are impossible being, respectively,
\[
\exp\left[\half it - \frac{1}{2x^2}\right]\left\{A\left(xe^{it}\right)^{(-1-i\sqrt{15}/2)}
 + B\left(xe^{it}\right)^{(-1+i\sqrt{15}/2)}\right\}
\]
and
\[
\exp\left[\half it - \frac{1}{2x^2}\right]\left\{A\left(xe^{it}\right)^{(-1-i\sqrt{3}/2)} + B\left(xe^{it}\right)^{(-1+i\sqrt{3}/2)}\right\},
\]
where $A$ and $B$ are constants of integration. Neither the normal
ordering method nor the Weyl quantisation procedure leads to a
result which is physical!

\section{The Last Multiplier of Jacobi}

Jacobi's Last Multiplier is a solution of the linear
partial differential equation \cite {Jacobi 86 a,Whittaker 44 a},
\begin {equation}
\sum_{i = 1} ^n \frac {\partial (Ma_i)} {\partial x_i} = 0.
\label{31.0}
\end {equation}
The relationship between the Jacobi Last Multiplier and the
Lagrangian, {\it videlicet}
\begin {equation}
\frac {\partial ^ 2L} {\partial\dot{x} ^ 2} = M \label {31.1}
\end {equation}
for a one-degree-of-freedom system, is perhaps not widely known.

If two multipliers, $M_1 $
and $M_2 $, are known, their ratio is a first integral.

In the
case of a conservative system with the standard energy integral
\begin {equation}
E = \half\dot {x} ^ 2+ V (x) \label {31.2}
\end {equation}
and Lagrangian
\begin {equation}
L = \half\dot {x} ^ 2- V (x) \label {31.3}
\end {equation}
it is evident from (\ref {31.1}) that one multiplier is a constant
-- taken to be 1 without loss of generality -- and so all
multipliers are first integrals.  This combined with (\ref {31.1})
is a simple recipe for the generation of a Lagrangian.  One has
\begin {equation}
\frac {\partial ^ 2L} {\partial\dot{x} ^ 2} =
1\quad\Longrightarrow\quad L = \half\dot {x} ^ 2+ \dot {x}f_1
(t,x) + f_2 (t,x), \label {31.4}
\end {equation}
where $f_1 $ and $f_2 $ are arbitrary functions of integration.
Naturally different multipliers give rise to different
Lagrangians.

Lagrange's equation of motion for (\ref
{31.4}) is
\begin {equation}
\ddot {x} + \frac {\partial f_1} {\partial t} - \frac {\partial
f_2} {\partial x} = 0 \label {31.5}
\end {equation}
whereas that for (\ref {31.3}) is
\begin {equation} \ddot {x} + V'(x) = 0.  \label {31.6}
\end {equation}
The requirement that the two Newtonian equations be the
same is
\begin {equation}
\frac {\partial f_1} {\partial t} - \frac {\partial f_2} {\partial
x} = V' (x ). \label {31.7}
\end {equation}
This constraint may be expressed through
setting
\begin {equation}
f_1 = \frac {\partial g} {\partial x},\quad f_2 = \frac {\partial
g} {\partial t} - V (x), \label {31.8}
\end {equation}
where $g (t,x) $ is an arbitrary function of its arguments.
Consequently the Lagrangian, (\ref {31.4}), becomes
\begin {equation}
L = \half\dot {x} ^ 2- V (x) + \dot {x}\displaystyle {\frac
{\partial g} {\partial x} +\frac {\partial g} {\partial t}} =
\half\dot {x} ^ 2- V (x) + \dot {g}, \label {31.9}
\end {equation}
{\it ie}, the functions $f_1 $ and $f_2 $ are a consequence of the
arbitrariness of a Lagrangian with respect to a total time
derivative, the gauge function.

\section {Algebraic Consistency of Gauge-Variant Lagrangians \cite{Nucci 07 a}}

The canonical momentum for (\ref {31.4}) is
\begin {equation}
p = \frac {\partial L} {\partial\dot {x}} = \dot {x} + f_1 \label
{32.1}
\end {equation}
so that
\begin {equation}
H  = p\dot {x} - L  =  \half p ^ 2- pf_1+ \half f_1 ^ 2- f_2
\label {32.2}
\end {equation}
is the Hamiltonian.  Whether one uses the Weyl quantisation
formula or the symmetrisation of $pf_1 $ makes no difference to the
form of the Schr\"odinger Equation corresponding to (\ref {32.2})
which is
\begin {equation}
2i\frac {\partial u} {\partial t} = -\frac {\partial ^ 2u}
{\partial x ^ 2} + 2 if_1\frac {\partial u} {\partial x} +\left
(f_1 ^ 2-2f_2+i\frac {\partial f_1} {\partial x}\right)u. \label
{32.3}
\end {equation}

The Schr\"odinger Equation, (\ref {32.3}), is quite general.  We
now introduce the simple harmonic oscillator with Newtonian
equation of motion $\ddot {x} +k ^ 2x = 0 $ so that the constraint
(\ref {31.7}) is
\begin {equation}
\frac {\partial f_1} {\partial t} - \frac {\partial f_2} {\partial
x} = k ^ 2x.  \label {32.4}
\end {equation}

The Lie point symmetries of (\ref {32.3}) subject to the constraint
(\ref {32.4}) are
\begin{eqnarray}
\Gamma_1 &=& \cos(k t)\,\partial_x+ \left[\cos(k t) f_1 - \sin(k t)  k  x\right] iu\,\partial_u   \nonumber\\
\Gamma_2 &=& - \sin(k t)\,\partial_x-\left[ \sin(k t) f_1+\cos(k t) k x\right ] i u\, \partial_u  \nonumber\\
\Gamma_3 &=& \partial_t+\left( f_2 +\frac{1}{2} \, k^2  x^2\right) iu\,\partial_u \nonumber\\
\Gamma_4 &=& \cos(2 k t)\,\partial_t -\sin(2 k t) k x\, \partial_x\nonumber\\
  &&+\left[i \cos(2 k t) \left( f_2 -\frac{1}{2}\, k^2 x^2\right)   - k\sin(2 k t)\left(  i   x f_1 -\frac{1}{2}\right)\right]u\partial_u   \nonumber\\
\Gamma_5 &=& - \sin(2 k t)\partial_t -\cos(2 k t) k x \partial_x\nonumber\\
  &&-\left[i\sin(2 k t) \left(f_2 -\frac{1}{2}\, k^2 x^2\right)   + k \cos(2 k t) \left( i  x f_1-\frac{1}{2}\right)  \right]u\partial_u   \nonumber\\
\Gamma_6 &=& u \,\partial_u \nonumber \\
\Gamma_7 & = & s (t,x)\partial_u, \label {32.5}
\end{eqnarray}
where $s (t,x) $ is a solution of (\ref {32.3}), which is a
representation of the well-known algebra, $\{sl (2,R)\oplus_s
W\}\oplus_s\infty A_1 $, of the Schr\"odinger Equation for the
one-dimensional linear oscillator and related systems.  The
presence of the functions $f_1 $ and $f_2 $ subject to the
constraint (\ref {32.4}) does not affect the number of Lie point
symmetries of (\ref {32.3}) vis-\`a-vis the number for the
Schr\"odinger Equation for the simple harmonic oscillator.

\section {Creation and Annihilation Operators}

We write $\Gamma_1 $ and $\Gamma_2 $ and
$\Gamma_4 $ and $\Gamma_5 $ as
\begin {eqnarray}
\Gamma_{1\pm} & = & \exp [\pm kit]\{\partial_x+i (f_1\pm i k x)u\partial_u\} \label {32.6} \\
\Gamma_{4\pm} & = & \exp [\pm 2 kit]\left\{\partial_t \pm
kix\partial_x+i\left [\left (f_2-\half k ^ 2x ^ 2\right) \pm
k\left (ixf_1-\half \right)\right]u\partial_u\right\}. \label
{32.7}
\end {eqnarray}
The normal route to the solution of the Schr\"odinger Equation,
(\ref {32.3}), is to use the symmetries (\ref {32.6}) which are the time-dependent
progenitors of the well-known creation and annihilation operators
of Dirac in the case that $f_1 $ and $f_2 $ are restricted as
above.

To solve the Schr\"odinger Equation, (\ref {32.3}), using Lie's
method we reduce (\ref {32.3}) to an ordinary differential equation
by using the invariants of the symmetries as a source of the
variables.  We must also be cognisant of the need for the solution
of (\ref {32.3}) to satisfy the boundary conditions at $\pm\infty
$.

With this requirement in mind we take $\Gamma_{1+} $.  The
associated Lagrange's system is
\begin {equation}
\frac {\d t} {0} = \frac {\d x} {1} = \frac {\d u} {i\left
(f_1+kix\right)u} \label {32.8}
\end {equation}
which gives the characteristics $t $ and $u\exp\left [\half kx ^
2-ig (t,x)\right] $, where we have made use of the first of (\ref
{31.8}) and the fact that $t $ is a characteristic.  To find the
solution corresponding to $\Gamma_{1+} $ we write
\begin {equation}
u (t,x) = h (t)\exp\left [-\half kx ^ 2+ig (t,x)\right], \label
{32.9}
\end {equation}
where $h (t) $ is to be determined, and substitute it into (\ref
{32.3}) which simplifies to
$$
i\dot{h} = \half kh
$$
so that
$$
h (t) = \exp\left [-\half kit\right]
$$
and
\begin {equation}
u (t,x) = \exp\left [-\half kit-\half kx ^ 2+ig (t,x)\right].
\label {32.10}
\end {equation}
With $g = 0 $ we
recognise the ground-state solution for the time-dependent
Schr\"odinger Equation of the simple harmonic oscillator.

We use $\Gamma_{1-} $ as a time-dependent `creation
operator'.  If we write the left hand side of (\ref {32.10}) as
$u_0 $, we can have a solution symmetry of the form
\begin {equation}
\Gamma_{70} = u_0 (t,x)\partial_u, \label {32.11}
\end {equation}
where the subscript, 7j, means that we are using the symmetry
$\Gamma_7 $ with the specific solution, $u_j (t,x) $. We use the
closure of the Lie algebra under the operation of taking the Lie
Bracket to obtain further solutions.  Thus
\begin {equation}
\left [\Gamma_{1-},\,\Gamma_{70}\right]_{LB}  =  \left\{-kx
+i\displaystyle {\frac {\partial g} {\partial x}} - (if_1
+kx)\right\}\exp\left [-\mbox {$\frac {3} {2} $} kit -\half kx ^
2+ig\right]\partial_u \label {32.12}
\end {equation}
so that we have
\begin {equation}
u_1 (t,x) = - 2kx \exp\left [-\mbox {$\frac {3} {2} $} kit -\half
kx ^ 2+ig\right]. \label {32.13}
\end {equation}
Likewise $\left [\Gamma_{1-},\,\Gamma_{71}\right]_{LB}  $ gives
\begin {equation}
u_2 (t,x) = \left (4k ^ 2x ^ 2- 2k \right)\exp\left [-\mbox
{$\frac {5} {2} $} kit -\half kx ^ 2+ig\right]. \label {32.14}
\end {equation}

$\Gamma_{4\pm} $ act as double annihilation and creation operators.

Finally the Lie Bracket of $i\Gamma_3 $ with $\Gamma_7 $ yields
the energy.  For example with $\Gamma_{72} $ one has
\begin {equation}
\left [i\Gamma_{3},\,\Gamma_{72}\right]_{LB}  =  \mbox {$\frac {5}
{2} $}ku_2\partial_u. \label {2.16}
\end {equation}

\section{A Proliferation of Lagrangians \cite{Nucci 07 b}}

Lie's method \cite {Lie 74 a,Lie 67 a} for the
calculation of the Jacobi Last Multiplier is firstly to find the
value of
\begin{equation}
\Delta = \mbox{\rm det}\left[\begin{array}{c} e_{ij}\\ s_{ij}
\end{array} \right], \label{41.1}
\end{equation}
in which the matrix is square with the elements $e_{ij}$ being the
vector field of the set of first-order differential equations by
which the system is described and the elements, $s_{ij}$, being
the coefficient functions of the number of symmetries of the given
system necessary to make the matrix square. If $\Delta$ is not
zero, the corresponding multiplier is $M=\Delta^{-1}$.

We use the
simple harmonic oscillator with equation of motion
\begin {equation}
\ddot {q} + k ^ 2q = 0 \label {41.01}
\end {equation}
as our vehicle.

To determine Jacobi's Last Multipliers one writes the system as a set of first-order ordinary differential equations and (\ref{41.01}) becomes
\begin {eqnarray}
& &\dot {u_1} = u_2 \nonumber \\
& &\dot {u_2} = -k ^ 2u_1 \label {410.1}
\end {eqnarray}
with associated vector field
\begin {equation}
X_{SHO} = \partial_t+ u_2\partial_{u_1} - k ^ 2u_1\partial_{u_2}.
\label {410.2}
\end {equation}

As a linear second-order ordinary differential equation (\ref
{41.01}) possesses eight Lie point symmetries.  In terms of the
variables used in (\ref {410.1}) the eight vectors are
\begin{equation}\label {410.3}\begin {array}{lcl}
\Gamma _{1} &=&  \cos kt{\partial_{u_1}} - k\sin kt{\partial_{u_2}}  \\
\Gamma _{2} &=& \sin kt{\partial_{u_1}} + k\cos kt{\partial_{u_2}}  \\
\Gamma _{3} &=& {u_1}{\partial_{u_1}} + {u_2}{\partial_{u_2}}  \\
\Gamma _{4} &=& \partial_t  \\
\Gamma _{5} &=& \cos 2kt\partial_t - k{u_1}\sin 2kt{\partial_{u_1}}
- \left(2k^{2}{u_1}\cos 2kt  - k{u_2}\sin 2kt\right){\partial_{u_2}} \\
\Gamma _{6} &=& \sin 2kt\partial_t + ku_{1}\cos 2kt{\partial_{u_1}}
 - \left(2k^{2}{u_1}\sin 2kt + k{u_2}\cos 2kt\right){\partial_{u_2}}  \\
\Gamma _{7} &=& {u_1}\cos kt\partial_t -k{u_1}^{2}\sin kt{\partial_{u_1}}
 - \left(k^{2}{u_1}^{2}\cos kt + k{u_1}{u_2}\cos kt + {u_2}^{2}\cos kt\right){\partial_{u_2}}
   \\
\Gamma _{8} &=& {u_1}\sin kt\partial_t +k{u_1}^{2}\cos
kt{\partial_{u_1}} - \left(k^{2}{u_1}^{2}\cos kt - k{u_1}{u_2}
\cos kt + {u_2}^{2}\sin kt\right){\partial_{u_2}}.
\end {array}\end{equation}

Since the vector field, (\ref {410.2}), has three elements, two symmetries of the eight in (\ref {410.3}) are required for the computation of the determinant.  There are twenty-eight possibilities.  Of these fourteen are zero.  Of the fourteen nonzero determinants there are really only three distinct possibilities.  The other multipliers can be expressed as combinations of these three.  Consequently we list only the three basic multipliers plus a single combination of obvious interest.  They are
\begin {eqnarray}
& & JLM_{12} = k \nonumber \\
& & JLM_{13} = \frac{1}{ku_1\sin k t+u_2\cos kt} \nonumber \\
& & JLM_{23} = \frac{1}{-ku_1\cos k t+u_2\sin kt} \nonumber \\
& & JLM_{34} = \left [JLM_{13}^{-2} + JLM_{23}^{-2}\right] ^ {- 1} =\frac{1}{u_2^2+k^2u_1^2}. \label {410.5}
\end {eqnarray}

For each of these four multipliers we can calculate a Lagrangian and we list them with the constraint imposed on the two functions of integration, $f_1(t,u_1)$ and $f_2(t,u_1)$, after the Lagrangian to which it applies.
 \begin {eqnarray}
 & &L_{12} = \half u_{2}^2+f_1u_{2}+f_2, \nonumber \\
& &\nonumber\\
& &\frac {\partial f_1}{\partial t} - \frac {\partial f_2}{\partial u_{1}} =  k^2u_{1}; \nonumber \\
& &\nonumber\end{eqnarray}
\begin {eqnarray} & &\hspace{-0.6cm}L_{13} =
\sec^2kt\left[\log (ku_{1}\sin kt  + u_{2}\cos kt
)\left(ku_{1}\sin kt  + u_{2}\cos kt \right) \right. \nonumber \\
&&
\left. -u_{2}\cos kt  - ku_{1}\sin kt\right] +f_1u_{2}+f_2,  \nonumber \\
& &\nonumber\\
& & {\frac {\partial f_1}{\partial t}} - {\frac {\partial f_2}{\partial u_{1}}} = 0; \nonumber \\
& &\nonumber\end{eqnarray}
\begin {eqnarray} & &\hspace{-1cm}L_{23} =
\cosec^2kt\left[\log (-ku_{1}\cos kt + u_{2}\sin kt
)\left(-ku_{1}\cos kt +u_{2}\sin kt \right)\right.
\nonumber \\ && \left. -u_{2}\sin kt  + ku_{1}\cos kt \right]+f_1u_{2}+f_2,  \nonumber \\
& &\nonumber\\
& & {\frac {\partial f_1}{\partial t}} - {\frac {\partial f_2}{\partial u_{1}}} = 0; \nonumber \\
& &\nonumber\end{eqnarray}
\begin{eqnarray} && L_{34}=
\frac{u_2}{ku_1}\,\arctan\left(\frac{u_2}{ku_1}\right)-\frac{1}{2}\,\log\left(\frac{u_2^2}{k^2u_1^2}+1\right)+f_1u_2+f_2,
 \nonumber\\& &\nonumber\\
& & u_{1} \left(\frac{\partial f_1}{\partial t}- \frac{\partial
f_2}{\partial u_{1}}\right) =
 1.\nonumber
\end{eqnarray}

The number of Noether point symmetries associated with these
Lagrangians varies \cite{Nucci 07 b}. There are five for $L_{12}
$, three for $L_{13} $ and $L_{23} $ and two for $L_{34} $, i.e.
\begin{eqnarray}
L_{12} &\Longrightarrow & \Gamma_1, \Gamma_2, \Gamma_4, \Gamma_5,
\Gamma_6 \nonumber\\
L_{13} &\Longrightarrow & \Gamma_1,
\Gamma_4+\Gamma_5,-k\Gamma_3+\Gamma_6 \nonumber\\
L_{23} &\Longrightarrow
&\Gamma_2,-\Gamma_4+\Gamma_5,k\Gamma_3+\Gamma_6 \nonumber\\
L_{34} &\Longrightarrow & \Gamma_3, \Gamma_4
\end{eqnarray}
$L_{12} $ is the only Lagrangian with five Noether point
symmetries.  Not one of the fourteen Lagrangians has four, an
additional three have three, seven more have two and there are
none with one or zero Noether point symmetries \cite{Nucci 07 b}.

For each of these Lagrangians one may construct an Hamiltonian.
They are, with the relationship between the momentum, $p$, and
$u_2$,
\begin {eqnarray*}
{u_2} & = & p-{f_1} \\
H_{12} & = & \half{p}^2-{p} {f_1}+\half {f_1}^2-{f_2}
\end{eqnarray*}
\begin{eqnarray*}
 {u_2} & = &  - ku_1\tan(k t) + \frac{\exp[\cos(k t)(p-f_1)]}{\cos(k t)} \\
H_{13} & = & \frac{\exp[\cos(k t)(p-f_1)]}{\cos^2(k t)}
-ku_1\tan(k t )(p-f_1)-f_2
\end{eqnarray*}
\begin{eqnarray*}
{u_2} & = &   ku_1\cot(k t) + \frac{\exp[\sin(k t)(p-f_1)]}{\sin(k t)} \\
H_{23} & = & \frac{\exp[\sin(k t)(p-f_1)]}{\sin^2(k t)}
+ku_1\cot(k t )(p-f_1)-f_2 \end{eqnarray*}
\begin{eqnarray*} {u_2} & = & k{u_1}{\tan}[k{u_1}(p-f_1)] \\
H_{34} & = & \half\log[\tan^2[ku_1(p-f_1)]+1]-f_2
\end {eqnarray*}
with the constraints on $f_1 $ and $f_2 $ listed above. Apart from
$H_{12} $ the determination of the corresponding Schr\"odinger
Equation is a nontrivial exercise.

\section{Conclusion}

If one accepts that a mathematical description of a physical
reality should be consistent with the Physics, it is quite evident
from the elementary examples we have considered here that the
standard approaches to the quantisation of a Classical Hamiltonian
System are fraught with the possibility of error.  In the case of
different representations of the one-dimensional simple harmonic
oscillator we have been able to present a consistent approach for
quantisation.  This is because of the generous supply of Lie point
symmetries with which this problem is endowed.  We have seen that
in addition to some of the standard examples presented in the
general literature that it is possible to construct Lagrangians
and hence Hamiltonians of a far greater number than one would
normally expect, indeed normally desire.  In our analysis of the
Noether point symmetries of these Lagrangians we saw that the
number of symmetries varied from two to five.  Given the close
connection of the Noether symmetries to the Lie symmetries of the
corresponding Schr\"odinger equation the question of the correct
choice of a Lagrangian can be quite important in the search for
closed-form solutions.  In addition we have the further question
of an appropriate method for quantisation.  By using the approach
which we advocated in \cite {Leach 06 a} we were able to obtain
the correct result for the Hamiltonian, (\ref {2.1}).  It has been
known for over fifty years \cite {van Hove 51 a}  that
quantisation and nonlinear canonical transformations have no
guarantee of consistency.  We argued then that there should be a
preservation of the algebraic structure.  {\it A fortiori} with
the plethora of Lagrangians for the standard representation of the
simple harmonic oscillator and the considerable variation in the
number of Noether symmetries the need for the preservation of the
algebraic structure becomes even more evident.

\section*{Acknowledgements}

PGLL thanks Professor MC Nucci and the Dipartimento di Matematica
e Informatica,
 Universit\`a di Perugia, for hospitality and the provision of facilities
  while this work was prepared and the University of KwaZulu-Natal for its continued support.

\end{document}